\documentclass[12pt]{article}
\usepackage{amsmath}
\usepackage{amssymb}
\newsymbol \blackbox 1004
\newcommand{\eh}{\hfill}\newlength{\sperr}

\newenvironment{proof}{{\settowidth{\sperr}{\bf\rm
Proof}%
\par\addvspace{0.3cm}\noindent\parbox[t]{1.3\sperr}
{\bf\rm P\eh r\eh o\eh o\eh f\eh }%
}}{\nopagebreak\mbox{} $\blackbox$\par\addvspace{0.3cm}}

\def\br{\breve}
\def\a{\alpha}
\def\b{\beta}
\def\g{\gamma}

\def\wh{\widehat}
\def\wt{\widetilde}
\def\ov{\overline}
\def\p{\partial}

\newtheorem{Pa}{Paper}[section]
\newtheorem{Tm}[Pa]{{\bf Theorem}}

\newtheorem{Rk}[Pa]{{\bf Remark}}

\newtheorem{Ee}[Pa]{{\bf Example}}

\newtheorem{Pn}[Pa]{{\bf Proposition}}
\newcommand{\CC}
{{\mathchoice {\setbox0=\hbox{$\displaystyle\rm C$}\hbox{\hbox
to0pt{\kern0.4\wd0\vrule height0.9\ht0\hss}\box0}}
{\setbox0=\hbox{$\textstyle\rm C$}\hbox{\hbox
to0pt{\kern0.4\wd0\vrule height0.9\ht0\hss}\box0}}
{\setbox0=\hbox{$\scriptstyle\rm C$}\hbox{\hbox
to0pt{\kern0.4\wd0\vrule height0.9\ht0\hss}\box0}}
{\setbox0=\hbox{$\scriptscriptstyle\rm C$}\hbox{\hbox
to0pt{\kern0.4\wd0\vrule height0.9\ht0\hss}\box0}}}}

\title{Nonisospectral integrable nonlinear  equations with  external potentials
and their GBDT solutions}
\author{Alexander Sakhnovich}

\date{}
\begin{document}
\maketitle
Fakult\"at f\"ur Mathematik, Universit\"at Wien, Nordbergstrasse
15, \\ A-1090 Wien, Austria

E-mail: al$_-$sakhnov@yahoo.com

\vspace{2em}
{\bf Short title:}  Nonisospectral  nonlinear  equations

\begin{abstract}
Auxiliary systems for matrix nonisospectral
equations, including coupled NLS with external potential
and KdV with variable coefficients, were introduced.
Explicit solutions of  nonisospectral
equations were constructed using the GBDT
version of the B\"acklund-Darboux transformation.
\end{abstract}


\section{Introduction} \label{intro}
\setcounter{equation}{0} The nonisospectral method to generate
solvable nonlinear evolution equations was proposed in the seminal
paper \cite{Cal} and actively used in a wide range of papers (see,
for instance, various references in the recent papers \cite{Ni,
Zha, Zho}). In particular, this approach allows to generate
physically interesting nonlinear equations with external
potentials \cite{CaDe1, CaDe1d}. Such equations are used to
describe deep water and plasma waves, waves in non-uniform media,
light pulses, and energy transport (see \cite{CSoS,  SuSu, Zha}
and references therein).

In the paper we  assume that the spectral parameter $\lambda$ depends
on both variables $x$ and $t$. In this way we construct solvable
generalizations of the matrix versions of the coupled nonlinear
Schr\"odinger (CNLS), KdV and MKdV equations. Matrix versions are
of interest (see, for instance, \cite{CaDe2}) and include, in
particular, scalar cases and  multi-component cases. We construct
auxiliary linear systems for our matrix (and scalar) cases and
some equations seem to be new in the scalar cases  too.

After that we apply to the constructed equations GBDT, which is a
version of the B\"acklund-Darboux transformation developed by the
author in \cite{SaA2}-\cite{SaZu}. The B\"acklund-Darboux
transformation is a well-known and fruitful tool to construct
explicit solutions of the integrable equations and some classical
linear equations as well. Some important versions of the
B\"acklund-Darboux transformation one can find in \cite{DeLo, Dei,
GeT, Gu, Mar, MS, ZM}. The GBDT version of the iterated
B\"acklund-Darboux transformation is of a rather general nature
and   provides simple algebraic formulas based on some system
theoretical and matrix identities results. The parameter matrices,
which are used in GBDT, have an arbitrary Jordan structure, while
diagonal matrices (with eigenvalues of the auxiliary spectral
problems as the entries) are mostly used in other approaches.

Examples are considered in a  more detailed way.

\section{Nonisospectral equations} \label{noniso}
\setcounter{equation}{0}
\subsection{NLS with external potential}
Recall the coupled nonlinear Schr\"odinger equation (CNLS) of the
form \cite{FO}:
\begin{equation} \label{cmNSE}
v_{1t}+iv_{1xx}+2iv_1v_2v_1=0, \quad \,
v_{2t}-iv_{2xx}-2iv_2v_1v_2=0 \quad (v_t:=\frac{\p}{\p t}v).
\end{equation}
We shall consider the matrix version of (\ref{cmNSE}), where
$v_1$ and $v_2$ are $m_1 \times m_2$ and $m_2 \times m_1$ ($m_1,
\, m_2 \, \geq 1$) matrix functions, respectively. Auxiliary
linear systems for CNLS are given by the formulae
\begin{equation} \label{bt13}
w_x(x,t, \lambda)=G(x,t, \lambda)w(x,t,\lambda), \quad w_t(x,t, \lambda)=F(x,t,
\lambda)w(x,t,\lambda).
\end{equation}
Here we have
\begin{equation} \label{1}
G=-(\lambda q_1+ q_0), \quad   F=-(\lambda^2 Q_2+ \lambda Q_1 +Q_0),
\end{equation}
where $\lambda$ is the independent of $x$ and $t$ spectral parameter,
\begin{equation} \label{2}
2q_{1}=-Q_{2}=2ij, \hspace{1em}2q_{0}(x,t)=-Q_{1}(x,t)=2j \xi
(x,t),
\end{equation}
\begin{equation} \label{3}
Q_{0}(x,t)= i(j \xi (x,t)^{2}- \xi_{x}(x,t)),
\end{equation}
\begin{equation}   \label{4}
j= \left[
\begin{array}{cc}
I_{m_1} & 0 \\ 0 & -I_{m_2}
\end{array}
\right], \hspace{1em} \xi = \left[\begin{array}{cc} 0&v_1
\\v_2&0\end{array}\right],
\end{equation}
and $I_k$ is the $k \times k $ identity matrix. We assume in this
section that $G$ and $F$  are continuous together with their first
derivatives. Then the compatibility condition for systems
(\ref{bt13}) can be written down in the zero curvature equation
form
\begin{equation} \label{5}
G_t(x,t,\lambda)-F_x(x,t, \lambda)+G(x,t,\lambda) F(x,t,\lambda)-F(x,t,\lambda)G(x,t,\lambda)=0.
\end{equation}
It can be checked directly that (\ref{cmNSE}) is equivalent to the
compatibility condition (\ref{5}). In other words, equation
(\ref{cmNSE}) can be presented in the zero curvature form
(\ref{5}). (See \cite{FT} on the historical details about zero
curvature  representation of the integrable equations.) It follows
from  (\ref{5}) that (\ref{cmNSE}) can be presented in the form:
\begin{equation} \label{6}
-j\Big(\xi_t(x,t)+ij\xi_{xx}(x,t)+2ij \xi(x,t)^3\Big)=0.
\end{equation}
Indeed, the left hand side in (\ref{6}) coincides with the left
hand side in (\ref{5}), while the coefficients at the degrees
$\lambda^k$ ($k>0$) on the left hand side in (\ref{5}) turn to zero. It
is easy to calculate that in the nonisospectral case (for $\lambda$
depending on $x$ and $t$) the zero curvature equation  (\ref{5}) is
equivalent to
\begin{equation} \label{7}
-\lambda_t q_1+2\lambda_x\lambda Q_2+\lambda_xQ_1-j\Big(\xi_t(x,t)+ij\xi_{xx}(x,t)+2ij
\xi(x,t)^3\Big)=0.
\end{equation}
Assume now
\begin{equation} \label{8}
\lambda_t=-4\lambda_x\lambda,
\end{equation}
to derive from (\ref{2}), (\ref{7}) and (\ref{8})
\begin{equation} \label{9}
\Big(\xi_t(x,t)+ij\xi_{xx}(x,t)+2ij \xi(x,t)^3\Big)+2\lambda_x
\xi(x,t)=0.
\end{equation}
Thus we obtain the following proposition
\begin{Pn}\label{Pn1}
The CNLS with external potential
\begin{equation} \label{10}
v_{1t}+iv_{1xx}+2iv_1v_2v_1+2\lambda_x v_1=0, \quad \,
v_{2t}-iv_{2xx}-2iv_2v_1v_2+2\lambda_x v_2=0
\end{equation}
is a nonisospectral integrable equation, which auxiliary linear
systems are given by (\ref{bt13})-(\ref{4}), where
$\lambda_t=-4\lambda_x\lambda$. In other words, equation (\ref{10}) admits zero
curvature representation (\ref{5}), where $G$ and $F$ are given by
(\ref{1})-(\ref{4}) and $\lambda_t=-4\lambda_x\lambda$.
\end{Pn}
\begin{Rk} \label{Rk3.2} The simplest solution of (\ref{9}) is given by the function
\begin{equation} \label{11}
\lambda(x,t)=\frac{1}{4}\big(x+c\big)\big(t+b\big)^{-1}.
\end{equation}
Correspondingly,  we obtain an integrable coupled nonlinear
Schr\"odinger equation (CNLS) with a simple external potential
\begin{equation} \label{0.1}
v_{1t}+iv_{1xx}+2iv_1v_2v_1+\frac{1}{2(t+b)} v_1=0, \quad \,
v_{2t}-iv_{2xx}-2iv_2v_1v_2+\frac{1}{2(t+b)}  v_2=0,
\end{equation}
\end{Rk}
The problem of similarity transformations \cite{MaPa, PKE} is of
interest here. When $b = \ov b$, using $\lambda(x,t)$ as in (\ref{11})
one can construct an integrable nonlinear Schr\"odinger equation
(NLS) with external potential \cite{SaAarx}, however the
substitution
\begin{equation} \label{0.1'}
t=-b-\wt t^{-1}, \quad x=-\wt x \wt t^{-1}
\end{equation}
turns it (see \cite{PKE}, Section 6, Example 4) into the classical
cubic NLS. In a quite similar way the substitution (\ref{0.1'})
and equalities
\begin{equation} \label{0.2}
\wt v_1=(t+b)\exp\Big(\frac{ix^2}{4(t+b)}\Big)v_1, \quad \wt
v_2=(t+b)\exp\Big(-\frac{ix^2}{4(t+b)}\Big)v_2
\end{equation}
transform (\ref{0.1}) in the case  $b = \ov b$ into the CNLS
\begin{equation} \label{0.3}
\wt v_{1 \wt t}+i \wt v_{1 \wt x \wt x}+2i\wt v_1 \wt v_2 \wt
v_1=0, \quad \, \wt v_{2 \wt t}-i\wt v_{2 \wt x \wt x}-2i\wt v_2
\wt v_1 \wt v_2=0.
\end{equation}
The case $b \not= \ov b$ is more interesting from that point of
view, though GBDT for the equation (\ref{0.1}) ($b=\ov b$) can be
applied to construct new solutions of (\ref{cmNSE}) and proves
therefore useful too.
\subsection{KdV equation with variable coefficients}
The matrix KdV equation can be written as:
\begin{equation} \label{K1}
v_{t} (x,t) -3v (x,t) v_{x} (x,t) -3v_{x} (x,t) v (x,t) +v_{xxx}
(x,t) =0,
\end{equation}
where $v$ is an $p \times p$ matrix function. Equation (\ref{K1})
admits zero curvature representation (\ref{5}), where polynomials
$G$ and $F$ of the form (\ref{1}) are defined via the $m \times m$
($m=2p$) coefficients
\begin{equation} \label{K2}
q_{1}= \left[ \begin{array}{lr}0 & 0
\\ I_{p} & 0 \end{array} \right], \hspace{1em}
q_{0}=- \left[ \begin{array}{lr}0 &  I_{p}
\\ v & 0 \end{array} \right],
\end{equation}
\begin{equation} \label{K3}
Q_{2}= \left[ \begin{array}{lr}0 & 0
\\ 4I_{p} & 0 \end{array} \right], \quad
Q_{1}=- \left[ \begin{array}{lr}0 &  4I_{p}
\\ 2v & 0 \end{array} \right], \quad
Q_{0}= \left[ \begin{array}{lr}v_{x} & -2v
\\ v_{xx} -2v^{2} & -v_{x} \end{array} \right].
\end{equation}
Now, substitute (\ref{K3}) by the equalities
\[
Q_{2}= g(t) \left[ \begin{array}{lr}0 & 0
\\ 4I_{p} & 0 \end{array} \right], \quad
Q_{1}=- g(t) \left[ \begin{array}{lr}0 &  4I_{p}
\\ 2v & 0 \end{array} \right],
\]
\begin{equation} \label{K4}
Q_{0}= g(t)\left(\left[ \begin{array}{lr}v_{x} & -2v
\\ v_{xx} -2v^{2} & -v_{x} \end{array} \right]+2f(t)j\right),
\quad j=\left[ \begin{array}{lr}I_p & 0 \\0 & -I_p \end{array}
\right],
\end{equation}
where $g$ and $f$ are scalar functions. One can easily take into
account corresponding changes in (\ref{5}) and obtain our next
proposition.
\begin{Pn}\label{Pn2}
a) Assume $\lambda=xf(t)+h(t)$. Then equation (\ref{5}), where $G$ and
$F$ are defined via (\ref{K2}) and (\ref{K4}), is a zero curvature
representation of KdV with variable coefficients and external
potential:
\begin{equation} \label{K5}
v_{t}+g(t)\big(v_{xxx}-3v  v_{x}  -3v_{x}  v -6fv)
=\big(x(f_t-12gf^2)+h_t-12fgh\big).
\end{equation}
b) Assume
\begin{equation} \label{K6}
\lambda_x=f, \quad \lambda_t=12fg\lambda, \quad f_t=12gf^2.
\end{equation}
Then equation (\ref{5}), where $G$ and $F$ are defined via
(\ref{K2}) and (\ref{K4}), is a zero curvature representation of a
special case of equation (\ref{K5}):
\begin{equation} \label{K7}
v_{t}+g(t)\big(v_{xxx}-3v  v_{x}  -3v_{x}  v -6fv) =0.
\end{equation}
\end{Pn}
Notice that equation (\ref{K5}), where $p=1$ (scalar case) and
$h=0$, was treated using the homogeneous balance principle in
\cite{Wa}. When (\ref{K6}) holds and $g=1$, one can put
$f=-(t+b)^{-1}/12$. The corresponding equation
\[
v_{t}+v_{xxx}-3v  v_{x}  -3v_{x}  v +\frac{1}{2(t+b)}v =0
\]
appeared  in \cite{Ba} and  its subcase $b=0$ is a well-known
cylindrical KdV \cite{CaDe1}.
\subsection{MKdV with external potential}
Introduce $G$ and $F$ by the equalities
\[
G=i\lambda j+\xi B, \quad F=-i \lambda^3 j -\lambda^2 \xi B-\frac{i\lambda}{2
}\Big(\xi_xBj+(\xi B)^2\Big)
\]
\begin{equation} \label{mk1}
+\frac{1}{4}\Big(\xi_{xx}B-(-1)^k(2\xi^3B+\xi_x\xi-\xi\xi_x)\Big)+
\frac{i}{2 }\p_x^{-1}\Big(\lambda_x(\xi B)^2\Big),
\end{equation}
where $j$ is given in (\ref{4}), $B=j^k$ ($k=0,1$), $\lambda=\ov
\lambda=\sqrt{(x+c)/3(t+b)   }$, and
\begin{equation} \label{mk2}
\xi= \left[ \begin{array}{lr}0 & -v^*
\\ v & 0 \end{array} \right].
\end{equation}
Then representation (\ref{5}) is equivalent to the equation
\[
4v_t=v_{xxx}+3(-1)^k(v_xv^*v+vv^*v_x)-\frac{4}{3(t+b)}v-2ifv_x
\]
\begin{equation} \label{mk3}
+2(-1)^ki\Big(v\p_x^{-1}\big(fv^*v\big)
-\p_x^{-1}\big(fvv^*\big)v\Big), \quad
f(x,t)=\Big(2\sqrt{3(x+c)(t+b)}\Big)^{-1}.
\end{equation}
Some other generalizations of MKdV one can find in \cite{Bi, Zha}.
\section{GBDT for the nonisospectral case: preliminaries} \label{Prel}
\setcounter{equation}{0} GBDT (nonisospectral case) for systems
with rational dependence on the spectral parameter have been
introduced in \cite{SaA4}. Here we shall need a reduction of the
theorem in \cite{SaA4} (Section 2, p. 1253) for the first order
systems of the form
\begin{equation} \label{bt1}
w^{\prime}(u, \lambda)={\mathcal G}(u, \lambda ) w(u, \lambda )
\quad (w^{\prime}=\frac{d}{du}), \hspace{1em} {\mathcal G}(u,
\lambda )=- \sum_{k=0}^{r} \lambda^{k}{\mathcal Q}_{k}(u).
\end{equation}
where the coefficients $ {\mathcal Q}_{k}(u)$ are $m \times m$
locally summable  on the interval $(-c_1,$ $ c_2)$   $\, (c_1, \,
c_2 \geq 0 )$ matrix functions. It was assumed in \cite{SaA4} that
the derivative of $\lambda=\lambda(u)$ rationally depends on $\lambda$, but now
it will suffice to suppose a polynomial dependence:
\begin{equation} \label{btn1}
\lambda^{\prime}(u)=\sum_{k=0}^r\omega_k(u)\lambda(u)^k.
\end{equation}
After fixing an integer $n>0$ the GBDT of the system (\ref{bt1})
is determined by the five parameter matrices: three $n \times n$
matrices $A_{1}(0)$, $A_{2}(0)$, and $S(0)$
  ($ \det S(0) \not= 0$)
and  two $n \times m$ matrices $\Pi_{1}(0)$ and $\Pi_{2}(0)$, such
that
\begin{equation} \label{bt2}
A_{1}(0)S(0)-S(0)A_{2}(0)= \Pi_{1}(0) \Pi_{2}(0)^{*}.
\end{equation}
Next, introduce matrix functions $A_{1}(u)$, $A_{2}(u)$,
$\Pi_{1}(u)$, $\Pi_{2}(u)$ and $S(u)$ with the given above values
at $u=0$ by the linear differential equations
\begin{equation} \label{btn2}
A_l^{\prime}(u)=\sum_{k=0}^r\omega_k(u)A_l(u)^k \quad (l=1,2),
\end{equation}
\begin{equation} \label{bt2'}
\Pi_{1}^{\prime}(u)= \sum _{k=0}^{r}A_{1}(u)^{k}
\Pi_{1}(u){\mathcal Q}_{k}(u), \quad \Pi^{\prime}_{2}(u)=-
\sum_{k=0}^{r}(A_{2}(u)^{*})^{k} \Pi_{2}(u){\mathcal
Q}_{k}(u)^{*},
\end{equation}
\begin{equation} \label{bt3}
S^{\prime}(u)=  \sum _{k=1}^{r} \sum _{j=1}^{k} A_{1}(u)^{k-j}
\Big(\Pi_{1}(u){\mathcal Q}_{k}(u) \Pi
_{2}(u)^{*}-\omega_k(u)S(u)\Big) A_{2}(u)^{j-1}.
\end{equation}
Notice that equations (\ref{btn2})-(\ref{bt3}) are chosen in such
a way that the identity
\begin{equation} \label{bt4}
A_{1}(u)S(u)-S(u)A_{2}(u)= \Pi_{1}(u) \Pi_{2}(u)^{*}
\end{equation}
follows from (\ref{bt2}) for all $u$ in the connected domain,
where  the coefficients ${\mathcal Q}_{k}$  are defined. (The
relation is obtained by the direct differentiation of   both sides
of (\ref{bt4}).) Assuming that $\det S(u) \not= 0$ and $\det
\Big(A_1(u)-\lambda(u)I_n\Big) \not= 0$ we  define a transfer matrix
function
\begin{equation} \label{bt5}
w_{A}(u, \lambda )=I_{m}- \Pi_{2}^{*}S^{-1}(A_{1}- \lambda
I_{n})^{-1} \Pi_{1}.
\end{equation}
\begin{Rk} \label{Rkbtn2}
Following the notation in the isospectral case we write
$w_A(u,\lambda)$, but as $\lambda=\lambda(u)$, so one can write simply $w_A(u)$.
\end{Rk}
Transfer matrix functions of the form
\[
w_A(\lambda)=I_{\mathcal L}- \Pi_2^*S^{-1}(A_1 - \lambda I_{\mathcal
H})^{-1} \Pi_1 \quad (A_1S-SA_2= \Pi_1 \Pi_2^*),
\]
where $I_{\mathcal L}$ and $I_{\mathcal H}$ are the identity matrices
in the Hilbert spaces  ${\mathcal L}$ and ${\mathcal H}$,
respectively, have been introduced and studied by L. Sakhnovich in
the context of his method of operator identities (see \cite{SaL1},
\cite{SaL3} and references in \cite{SaL3}) and take roots in the
M.~S. Liv\v{s}ic characteristic matrix functions. Our next theorem
is a reduction of the theorem from \cite{SaA4}.
\begin{Tm} \label{Tmbtn}
Let relations (\ref{btn1}) and (\ref{bt2}) hold.  Define matrix
functions $A_{1}$, $A_{2}$, $\Pi_1$, $\Pi_2$ and $S$ by the
equalities (\ref{btn2})-(\ref{bt3}). Then in the points of
invertibility of the matrix functions $S(u)$ and
$A_1(u)-\lambda(u)I_n$ the equation
\begin{equation} \label{bt8}
w_A^{\prime}(u, \lambda)= \wh{\mathcal G}(u, \lambda ) w_A(u, \lambda )
-w_A(u, \lambda){\mathcal G}(u, \lambda), \quad
  \wh{\mathcal G}(u, \lambda )=- \sum_{k=0}^{r}
\lambda(u)^{k} \wh{\mathcal Q}_{k}(u)
\end{equation}
is true, and the coefficients $\wh{\mathcal Q}_k$ are given by the
formulas
\begin{eqnarray} \label{btn8}
&& \wh{\mathcal Q}_{k} ={\mathcal Q}_{k} -
\sum_{j=k+1}^{r}\Big({\mathcal Q}_{j} Y_{j-k-1} -X_{j-k-1}
{\mathcal Q}_{j} + \nonumber
\\
&& + \sum _{s=k+2}^{j} X_{j-s} {\mathcal Q}_{j} Y_{s-k-2}\Big)
+\sum_{j=k+2}^{r}\omega_j\sum_{s=k+2}^{j} Z_{j-s,s-k-2},
\end{eqnarray}
where
\begin{equation} \label{bt10}
X_{k} = \Pi_{2}^{*}S^{-1}A_{1}^{k} \Pi_{1}, \hspace{2em} Y_{k} =
\Pi_{2}^{*}A_{2}^{k}S^{-1} \Pi_{1},
\end{equation}
\begin{equation} \label{btn3}
Z_{k,j} = \Pi_{2} ^{*}S^{-1}A_{1}^{k}SA_2^j S^{-1}\Pi_{1}.
\end{equation}
\end{Tm}
\begin{Rk} \label{Rkbtn3}
According to Theorem \ref{Tmbtn} the matrix function $w_A$ is a
Darboux matrix, which transforms solution $w$ of system
(\ref{bt1}) into solution $\wh{w}=w_Aw$ of the system
$\wh{w}^{\prime}= \wh{\mathcal G} \wh{w}$.
\end{Rk}
Finally, we shall need also formula (9) from \cite{SaA4}:
\begin{equation} \label{btn4}
\big(\Pi_{2}^*S^{-1}\big)^{\prime}=- \sum_{k=0}^{r}\br{\mathcal
Q}_{k}\Pi_{2}^*S^{-1}A_1^k, \quad \br{\mathcal
Q}_{k}:=\wh{\mathcal Q}_{k}-(k+1)\omega_{k+1} I_m \quad (\omega_{r+1}=0).
\end{equation}
\section{Solutions of equations with external potentials
} \label{sec3} \setcounter{equation}{0} In this section we shall
construct GBDT solutions for the CNLS equation (\ref{0.1}) and for
the KdV type equation (\ref{K7}).
First,  let  $v_1$ and $v_2$ satisfy (\ref{0.1}). It means that
zero curvature equation (\ref{5}), where $G$ and $F$ are defined
by (\ref{1})-(\ref{4}) and
\begin{equation} \label{12}
\lambda_x=f(t), \quad \lambda_t=-4f(t)\lambda \quad (f(t)=\frac{1}{4}(t+b)^{-1}),
\end{equation}
holds. Now, put $m=m_1+m_2$ and apply  GBDT from Section
\ref{Prel} for systems (\ref{bt13}). Putting $u=x$ and using the
notations of (\ref{btn1}), we derive from (\ref{1}) and (\ref{12})
that $r=1$, $\omega_1=0$, $\omega_0(x)\equiv {\mathrm{const}}$, i.e.,
$\omega_0(x,t)=f(t)$, where $t$ is a second variable. Putting $u=t$,
we derive $r=2$, $\omega_2=\omega_0=0$, $\omega_1(t)=-4f(t)$. Hence, formula
(\ref{btn2}) for $A=A_1, \, A_2$ after substitutions $u=x$ and
$u=t$ takes the form
\begin{equation} \label{13}
A_x=f(t)I_n, \quad A_t=-4f(t)A .
\end{equation}
Therefore we can put
\begin{equation} \label{14}
A_1(x,t)=f(t)\big(xI_n+a_1\big), \quad
A_2(x,t)=f(t)\big(xI_n+a_2\big),
\end{equation}
where $a_1$ and $a_2$ are some arbitrary $n \times n $ matrices.
We require (compare with (\ref{bt2})):
\begin{equation} \label{vst}
A_1(0,0)S(0,0)-S(0,0)A_2(0,0)=\Pi_1(0,0)\Pi_2(0,0)^*.
\end{equation}
Next, we introduce matrix functions $\Pi_1(x,t)$ and $\Pi_2(x,t)$,
where the dependence on $x$ is determined by the system $w_x=Gw$,
that is, by the coefficients of $G$, and the dependence on $t$ is
determined by the system $w_t=Fw$. Namely, when we put $u=x$, the
system for $\Pi_1$ in (\ref{bt2'}) takes the form
\begin{equation} \label{15}
\big(\Pi_1(x,t)\big)_x=A_1(x,t)\Pi_1(x,t)q_1+\Pi_1(x,t)q_0(x,t),
\end{equation}
and when we put $u=t$, the system takes the form
\begin{equation} \label{16}
\big(\Pi_1(x,t)\big)_t=A_1(x,t)^2\Pi_1(x,t)Q_2+A_1(x,t)\Pi_1(x,t)Q_1(x,t)+\Pi_1(x,t)Q_0(x,t).
\end{equation}
The compatibility of systems  (\ref{15}) and (\ref{16}) follows
from (\ref{5}). In a similar way we rewrite the second equation in
(\ref{bt2'}):
\begin{equation} \label{17}
\big(\Pi_2\big)_x= -\sum_{k=0}^1\big(A_2^*\big)^k\Pi_2 q_k^*,
\quad \big(\Pi_2\big)_t= -\sum_{k=0}^2\big(A_2^*\big)^k\Pi_2
Q_k^*.
\end{equation}
Recall now relations
\begin{equation} \label{18}
\omega_1=0, \quad \omega_0=f(t), \, \mathrm{when} \,u=x; \quad
\omega_2=\omega_0=0, \quad \omega_1=-4f(t), \, \mathrm{when} \,u=t.
\end{equation}
In view of (\ref{18}) we rewrite (\ref{bt3}) as
\begin{equation} \label{19}
S_x=\Pi_1q_1\Pi_2^*, \quad
S_t=\Pi_1Q_1\Pi_2^*+A_1\Pi_1Q_2\Pi_2^*+\Pi_1Q_2\Pi_2^*A_2+4f(t)S.
\end{equation}
According to (\ref{14})-(\ref{17}) and (\ref{19})  the matrix identity
\begin{equation} \label{20}
A_1S-SA_{2}\equiv  \Pi_{1} \Pi_{2}^{*}
\end{equation}
is true.  Finally, using (\ref{btn8}))
and equality $X_0=Y_0$, define coefficients $\wh q_k(x,t)$ and
$\wh Q_k(x,t)$:
\begin{equation} \label{21}
\wh q_1=q_1, \quad \wh q_0=q_0+X_0q_1-q_1X_0, \quad \wh Q_2=Q_2,
\quad \wh Q_1=Q_1+X_0Q_2-Q_2X_0,
\end{equation}
\begin{equation} \label{22}
\wh Q_0=Q_0+X_0Q_1-Q_1X_0+X_1Q_2-Q_2Y_1-X_0Q_2X_0.
\end{equation}
Partition matrix functions $\Pi_l$ ($l=1,2$): $\Pi_1=[\Lambda_1 \quad
\Lambda_2]$ and $\Pi_2=[\Psi_1 \quad \Psi_2]$, where $\Lambda_1$,
$\Psi_1$ are $n\times m_1$ blocks, and $\Lambda_2$, $\Psi_2$ are
$n\times m_2$ blocks. From Theorem \ref{Tmbtn} and Remark
\ref{Rkbtn3} follows proposition.
\begin{Pn}\label{PnNSEEP} Let  $v_1$ and $v_2$ satisfy CNLS
(\ref{0.1}) with external potential. Then, in the points of invertibility
of $S$ and $A_1-\lambda I_n$, the matrix functions
\begin{equation} \label{23}
\wh v_1=v_1-2i \Psi_1^*S^{-1}\Lambda_2, \quad \wh v_2=v_2-2i
\Psi_2^*S^{-1}\Lambda_1
\end{equation}
satisfy equation (\ref{0.1}) too.
\end{Pn}
\begin{proof}.
Equations (\ref{bt5}), (\ref{14})-(\ref{17}) and (\ref{19}) define
the transfer matrix function $w_A(x,t,\lambda)$. By Remark \ref{Rkbtn3}
for $\wh w=w_Aw$ we have  ${\wh w}_x=\wh G \wh w$ and ${\wh
w}_t=\wh F \wh w$, and so these two systems are compatible. That
is, the compatibility condition
\begin{equation} \label{24}
{\wh G}_t-{\wh F}_x+ \wh G \wh F-\wh F \wh G=0
\end{equation}
holds. Recall that formulas (\ref{1})-(\ref{4}) imply the
equivalence of the compatibility condition (\ref{5}) to the
equation (\ref{0.1}). Taking into account (\ref{23}), put
\begin{equation} \label{25}
\wh \xi = \left[\begin{array}{cc} 0& \wh v_1
\\ \wh v_2&0\end{array}\right]=\xi +i(jX_0j-X_0).
\end{equation}
When the equalities (\ref{2}) and (\ref{3}) remain valid after
substitution of the matrix functions $\{q_k\}, \, \{Q_k\}$ and
$\xi$ by the matrix functions $\{\wh q_k\}, \, \{\wh Q_k\}$ and
$\wh \xi$, respectively, then $\wh G $ and $\wh F$ have the same
structure as $G $ and $ F$. Therefore, similar to (\ref{5})
equation (\ref{24}) is equivalent to (\ref{0.1}), that is, $\wh
v_1$ and $\wh v_2$ satisfy (\ref{0.1}). It is easy to see that
according to (\ref{btn8}) we have
\begin{equation} \label{26}
\wh q_1=q_1, \quad \wh Q_2=Q_2.
\end{equation}
From (\ref{btn8}) it follows also that
\begin{equation} \label{27}
2 \wh q_0=-\wh
Q_1=2(j\xi-ijX_0+iX_0j)=2j\big(\xi+i(jX_0j-X_0)\big).
\end{equation}
Substitute (\ref{25}) into (\ref{27}) to get the relation similar
to the second relation in (\ref{2}):
\begin{equation} \label{28}
2 \wh q_0=-\wh Q_1=2j \wh \xi .
\end{equation}
It remains to prove that
\begin{equation} \label{29}
\wh Q_{0}(x,t)= i(j \wh {\xi} (x,t)^{2}- \wh {\xi}_{x}(x,t)).
\end{equation}
Recall that by the definition (\ref{bt10}) the equality $X_0=Y_0$
is true. Hence, by (\ref{2}) and (\ref{btn8})   we obtain
\begin{equation} \label{30}
\wh Q_{0}= i(j\xi^2-\xi_x)+2j \xi X_0- 2X_0 j \xi+2ij
Y_1-2iX_1j+2iX_0jX_0.
\end{equation}
According to  (\ref{20}) we have
$A_2S^{-1}=S^{-1}A_1-S^{-1}\Pi_1\Pi_2^*S^{-1}$. Therefore, in view
of (\ref{bt10}), it follows that
\begin{equation} \label{31}
Y_1=\Pi_2^*(S^{-1}A_1-S^{-1}\Pi_1\Pi_2^*S^{-1})\Pi_1=X_1-X_0^2.
\end{equation}
Using (\ref{31}), we rewrite (\ref{30}) as
\begin{equation} \label{32}
\wh Q_{0}= i(j\xi^2-\xi_x)+2i(jX_1-X_1j)-2ijX_0^2+2(j \xi X_0- X_0
j \xi)+2iX_0jX_0.
\end{equation}
To calculate  $-i \wh {\xi}_{x}$ notice that by (\ref{bt10}),
(\ref{btn4}), (\ref{15}) and (\ref{18}) we have
\begin{equation} \label{33}
\big(X_0\big)_x=-\wh q_1 X_1-\wh q_0 X_0+X_1q_1+X_0q_0.
\end{equation}
Taking into account (\ref{2}), (\ref{26}) and (\ref{28}), we
rewrite (\ref{33}) as
\begin{equation} \label{34}
\big(X_0\big)_x=-i j X_1- j \wh \xi X_0+i X_1 j+X_0 j \xi.
\end{equation}
By (\ref{25}) and (\ref{34}) we obtain
\begin{equation} \label{35}
-i \wh {\xi}_{x}=-i \xi_x+2i(jX_1-X_1j)-\wh {\xi}X_0j+j\wh
{\xi}X_0+jX_0j\xi j-X_0 j \xi.
\end{equation}
Finally, in view of (\ref{25}), (\ref{32}), (\ref{35}) and
equality $\xi j=-j \xi$, direct calculation shows that
\begin{equation} \label{36}
\wh Q_{0}+i\wh {\xi}_{x}=i j \wh {\xi}^2.
\end{equation}
Thus, equality (\ref{29}) follows.
\end{proof}
\begin{Rk}
Notice that for the trivial initial solution, that is, for the case
$v_1=0$ and $v_2=0$ the blocks of the matrix functions $\Pi_1$ and $\Pi_2$
are calculated via (\ref{14})-(\ref{17}) explicitly:
\begin{equation} \label{37}
\Lambda_s=\Big(\exp (-1)^{s+1}c_1(x,t)\Big)h_s, \quad c_1(x,t)=\frac{i}{2}f(t)(xI_n+a_1)^2,
\end{equation}
\begin{equation} \label{38}
\Psi_s=\Big(\exp (-1)^{s+1}c_2(x,t)\Big)k_s, \quad c_2(x,t)=\frac{i}{2}\ov{f(t)}(xI_n+a_2^*)^2,
\end{equation}
where  $h_s$ and $k_s$ are $n \times m_s$ matrices ($s=1,2$).
In this way, using (\ref{23}), a wide class of solutions of the matrix 
CNLS with external potential can be constructed explicitly.
\end{Rk}
\begin{Ee}
Assume $n=1$ (i.e., $A_1$ and $A_2$ are now scalar functions) and $a_1\not=a_2$. Then  the recovery of $S(x,t)$
is especially simple. Using (\ref{20}), we obtain
\begin{equation} \label{39}
S(x,t)=\Big((A_1(x,t)-A_2(x,t)\Big)^{-1}\Big(\Lambda_1(x,t)\Psi_1(x,t)^*+
\Lambda_2(x,t)\Psi_2(x,t)^*\Big).
\end{equation}
By (\ref{14}), (\ref{37}) and (\ref{38}) we rewrite (\ref{39}) as:
\[
S(x,t)=\frac{1}{f(t)(a_1-a_2)}\Big(\exp\big\{\frac{i}{2}f(t)\big(2x(a_1-a_2)+a_1^2-a_2^2\big) \big\}h_1k_1^*\Big.
\]
\begin{equation} \label{40}
+\exp\big\{\frac{i}{2}f(t)\big(2x(a_2-a_1)+a_2^2-a_1^2\big) \big\}h_2k_2^*\Big).
\end{equation}
From (\ref{23}), (\ref{37}) and (\ref{38}) it follows that
\begin{equation} \label{41}
\wh v_1=-\frac{2i}{S(x,t)}
\exp\{-\frac{i}{2}f(t)\big((x+a_1)^2+(x+a_2)^2\big) \}k_1^*h_2,
\end{equation}
\begin{equation} \label{42}
\wh v_2=-\frac{2i}{S(x,t)}
\exp\{\frac{i}{2}f(t)\big((x+a_1)^2+(x+a_2)^2\big) \}k_2^*h_1,
\end{equation}
where $S$ is given by 
(\ref{40}).
\end{Ee}

Now, let us construct GBDT for the KdV type equation (\ref{K7}).
We shall construct self-adjoint solutions. For simplicity we shall
put also the initial solution $v=0$. (This assumption is often
made, when explicit solutions are constructed.) Partition
$\Pi_1(x,t)$ into two $p \times p$ blocks: $\Pi_1=[\Lambda_1 \quad
\Lambda_2]$. The coefficients in (\ref{bt2'}) and  (\ref{bt3}) for
the cases $u=x$ and $u=t$ are given by the equalities (\ref{K2})
and (\ref{K4}), respectively, after substitution $v=0$. In other
words, the first relation in (\ref{bt2'}) can be rewritten as
\begin{equation} \label{K8}
\frac{ \partial}{ \partial x} \Lambda_{1}= \alpha  \Lambda_{2},
\hspace{1em} \frac{ \partial }{ \partial x} \Lambda_{2}= -
\Lambda_{1},
\end{equation}
\begin{equation} \label{K8'}
\frac{ \partial }{ \partial t}\Lambda_{1}=g\big(4 \alpha^{2}
\Lambda_{2}+2f\Lambda_{1}\big), \hspace{1em} \frac{ \partial }{
\partial t}\Lambda_{2}=-g\big(4 \alpha  \Lambda_{1}+2f
\Lambda_{2}\big),
\end{equation}
where $\a=A_1$. According to (\ref{K6}) formula (\ref{btn2})  ($l=1$)
takes the form
\begin{equation} \label{Kd1}
\a_x=fI_n, \quad \a_t=12fg\a.
\end{equation}
Suppose
\begin{equation} \label{K8''}
g(t)=\ov{g(t)}, \quad  f(t)=\ov{f(t)}.
\end{equation}
Taking into account (\ref{K8})-(\ref{K8''})  it is easy to check,
that $A_2$ and $\Pi_2$ given by
\begin{equation} \label{K9}
\Pi_2=\Pi_1 J^*, \quad  A_2=\a^*, \quad J=\left[
\begin{array}{cc}
0 & I_p \\ -I_p & 0
\end{array}
\right],
\end{equation}
satisfy relation
(\ref{btn2}) ($l=2$) and the second relation in (\ref{bt2'}) for $u=x$ and $u=t$.
In view of (\ref{K6}) relations (\ref{bt3})    take the form
\begin{equation} \label{K10}
S_x= \Lambda_{2} \Lambda_{2}^{*}, \quad S_t=4g\big( \alpha
\Lambda_{2} \Lambda_{2}^{*}+ \Lambda_{2} \Lambda_{2}^{*}
\alpha^{*}+ \Lambda_{1} \Lambda_{1}^{*}\big)-12fgS.
\end{equation}
Finally, relations  (\ref{btn4}) take the form
\begin{equation} \label{K11}
\big(\Pi_2^*S^{-1}\big)_x= - \sum_{k=0}^{1}\wh
q_{k}\Pi_{2}^*S^{-1}\a^k, \quad \big(\Pi_2^*S^{-1}\big)_t=
12fg\Pi_2^*S^{-1} - \sum_{k=0}^{2}\wh Q_{k}\Pi_{2}^*S^{-1}\a^k,
\end{equation}
where coefficients $\wh q_k$ and $\wh Q_k$ are obtained via
transformation (\ref{btn8}). Equality (\ref{bt2}) can be written
as
\begin{equation} \label{K12}
\a(0,0) S(0,0)-S(0,0)\a(0,0)^*=\Lambda(0,0)J\Lambda(0,0)^*.
\end{equation}
It yields the identity
\begin{equation} \label{K12'}
\a(x,t) S(x,t)-S(x,t)\a(x,t)^*=\Lambda(x,t)J\Lambda(x,t)^*.
\end{equation}
\begin{Pn}\label{PnKdVEP} Let  relations (\ref{K6}), (\ref{K8})-(\ref{K8''}), (\ref{K10})
and (\ref{K12}) hold, and let $\det S\not=0$. Put
\begin{equation} \label{K13}
\wh v=2(\Omega_{12}+\Omega_{21}+\Omega_{22}^2), \quad
\Omega_{kj}:=\Lambda_k^*S^{-1}\Lambda_j.
\end{equation}
Then, in the points of invertibility
of $S$, the matrix function $\wh v$ is 
a self-adjoint solution of the KdV type equation
(\ref{K7}).
\end{Pn}
\begin{proof}.
Similar to the isospectral KdV case,  the transformation (\ref{btn8})
of the coefficients $ q_k$ and $Q_k$ does not preserve their
structure. Therefore we cannot use Theorem \ref{Tmbtn} in this
proof, and so we will show that $\wh v$ satisfies  (\ref{K7})
directly, using (\ref{btn4}), i.e., formulae (\ref{K11}).
The GBDT solution of the isospectral classical KdV equation was
treated in detail in \cite{GKS}, Section 5. Relations (\ref{K8})
and the first relation in (\ref{K10}) coincide with formula (5.6)
in \cite{GKS}. By these relations we get
\begin{equation}\label{K14}
\frac{ \partial  }{ \partial
x}\Omega_{12}=-\Omega_{12}\Omega_{22}-\Omega_{11}+ \Lambda_{2}^{*}
\alpha^{*}S^{-1} \Lambda_{2}, \hspace{1em} \frac{
\partial  }{ \partial x}\Omega_{21}=-\Omega_{22}\Omega_{21}-\Omega_{11}+
\Lambda_{2}^{*}S^{-1} \alpha \Lambda_{2},
\end{equation}
\begin{equation}\label{K15}
\frac{ \partial }{ \partial x}\Omega_{22} =- (\Omega_{22}^{2}+
\Omega_{12}+ \Omega_{21})=- \frac{1}{2}\wh v, \hspace{1em} \frac{
\partial
 }{
\partial x}\Omega_{22}^{2}=- \frac{1}{2}(\wh v \Omega_{22}+ \Omega_{22}\wh v),
\end{equation}
\begin{equation}\label{K16}
\frac{ \partial  }{ \partial
x}\Omega_{11}=-\Omega_{12}\Omega_{21}+ \Lambda_{2}^{*}
\alpha^{*}S^{-1} \Lambda_{1}+ \Lambda_{1}^{*}S^{-1} \alpha
\Lambda_{2}.
\end{equation}
After proper change of notation, definition (\ref{K13}) coincides
with formula (5.16) in \cite{GKS}. Using relations (\ref{K8}),
(\ref{K11}) and (\ref{K12'})-(\ref{K16}) similar to \cite{GKS} we
obtain derivatives $\wh v_{xx}$ and $\wh v_{xxx}$. Some additional
terms appear as $\a$ depends now on $x$, i.e., $\a_x=f(t)I_p$,
whereas we had $\a\equiv $ const  in the isospectral case \cite{GKS}.
 In
particular, we have
\[
3\wh v^{2}-  \wh v_{xx}= 8(\Omega_{21}\Omega_{12}+
\Lambda_{2}^{*}S^{-1} \alpha \Lambda_{2} \Omega_{22}+\Omega_{22}
\Lambda_{2}^{*}
  \alpha^{*}S^{-1} \Lambda_{2}
\]
\begin{equation}\label{K17}
+\Lambda_{2}^{*}S^{-1} \alpha \Lambda_{1}+ \Lambda_{1}^{*}
  \alpha^{*}S^{-1} \Lambda_{2})-4f\Omega_{22},
\end{equation}
where $-4f\Omega_{22}$ is such an additional term (compare with
formula (5.31) in \cite{GKS}). After differentiation of the right
hand side of (\ref{K17}) we obtain
\begin{equation}\label{K18}
\big(\wh v^{2}- \wh v_{xx}\big)_x=R_1+R_2,
\end{equation}
where
\[
R_1=8 \Big( \Lambda_{2}^{*} ( \alpha^{*})^{2}S^{-1} \Lambda_{2}+
\Lambda_{2}^{*}S^{-1} \alpha ^{2} \Lambda_{2}-( \Lambda_{1}^{*}
  S^{-1} \alpha \Lambda_{1}+ \Lambda_{1}^{*}
\alpha^{*}S^{-1} \Lambda_{1} +( \Lambda_{1}^{*}S^{-1} \alpha
\Lambda_{2}\]\[ + \Lambda_{1}^{*}
  \alpha^{*}S^{-1} \Lambda_{2}
+ \Lambda_{2}^{*}S^{-1} \alpha  \Lambda_{1} ) \Omega_{22}+
\Lambda_{2}^{*}S^{-1} \alpha  \Lambda_{2}\Omega_{21}+\Omega_{12}
\Lambda_{2}^{*}
  \alpha^{*}S^{-1} \Lambda_{2}
\]\[+\Omega_{11}\Omega_{12}+\Omega_{21}\Omega_{11}+ \Omega_{22}( \Lambda_{2}^{*}
  \alpha^{*}S^{-1} \Lambda_{1}
+ \Lambda_{1}^{*}
  \alpha^{*}S^{-1} \Lambda_{2} \]\[
+ \Lambda_{2}^{*}S^{-1} \alpha  \Lambda_{1} )+\Omega_{22}(
\Lambda_{2}^{*}
  S^{-1} \alpha \Lambda_{2}
\Omega_{22}+\Omega_{22} \Lambda_{2}^{*}
  \alpha^{*}S^{-1} \Lambda_{2}
+\Omega_{21}\Omega_{12})  \]
\begin{equation}\label{K19}
+( \Lambda_{2}^{*}
  S^{-1} \alpha \Lambda_{2}
\Omega_{22}+ \Omega_{22} \Lambda_{2}^{*}
  \alpha^{*}S^{-1} \Lambda_{2}
+\Omega_{21}\Omega_{12}) \Omega_{22}) \Big),
\end{equation}
and $R_2$ is the additional, with respect to the isospectral case,
term:
\[
R_2=16f\Omega_{22}^2+8f(\Omega_{12}+\Omega_{21})-4f\big(\Omega_{22}\big)_x.
\]
In view of (\ref{K13}) and (\ref{K15}) we have
\begin{equation}\label{K20}
R_2= 8f\Omega_{22}^2+6f\wh v.
\end{equation}
Now, consider $\wh v_t$. By (\ref{K6}), (\ref{btn8}) and
(\ref{K9}) we get
\[
\wh Q_2=Q_2, \quad \wh
Q_1=Q_1-Q_2J\Pi_1^*S^{-1}\Pi_1+J\Pi_1^*S^{-1}\Pi_1Q_2,
\]
\[
\wh Q_0=Q_0-Q_1J\Pi_1^*S^{-1}\Pi_1+J\Pi_1^*S^{-1}\Pi_1Q_1
\]
\begin{equation}\label{K21}
-Q_2J\Pi_1^*\a^*S^{-1}\Pi_1+J\Pi_1^*S^{-1}\a\Pi_1Q_2-J\Pi_1^*S^{-1}\Pi_1Q_2J\Pi_1^*S^{-1}\Pi_1.
\end{equation}
From (\ref{K8'}), the second equality in (\ref{K11}), (\ref{K21})
and identity (\ref{K12'}) it follows that
\[
\frac{ \partial  }{ \partial t}\Omega_{12}=4 g \Big(
\Lambda_{2}^{*} ( \alpha^{*})^{2}S^{-1} \Lambda_{2}-
\Lambda_{1}^{*}S^{-1} ( \alpha \Lambda_{2} \Lambda_{2}^{*}+
\Lambda_{2} \Lambda_{2}^{*} \alpha^{*}+ \Lambda_{1}
\Lambda_{1}^{*}) S^{-1} \Lambda_{2}
\]
\begin{equation}\label{K22}
-
\Lambda_{1}^{*}
  S^{-1} \alpha \Lambda_{1} \Big)+12gf\Omega_{12} ,
\end{equation}
\[
\frac{ \partial }{ \partial t} \Omega_{21}=4g \Big(
\Lambda_{2}^{*}S^{-1} \alpha ^{2} \Lambda_{2}-
\Lambda_{2}^{*}S^{-1} ( \alpha \Lambda_{2} \Lambda_{2}^{*}+
\Lambda_{2} \Lambda_{2}^{*} \alpha^{*}+ \Lambda_{1}
\Lambda_{1}^{*}) S^{-1} \Lambda_{1}
\]
\begin{equation}\label{K23}
-
\Lambda_{1}^{*} \alpha^{*}S^{-1} \Lambda_{1} \Big)+12gf\Omega_{21}
,
\end{equation}
\[
\frac{ \partial  }{ \partial t}\Omega_{22}=-4g \Big(
\Lambda_{1}^{*}
  \alpha^{*}S^{-1} \Lambda_{2}+
\Lambda_{2}^{*}S^{-1} ( \alpha  \Lambda_{2} \Lambda_{2}^{*}+
\Lambda_{2} \Lambda_{2}^{*} \alpha^{*}+ \Lambda_{1}
\Lambda_{1}^{*}) S^{-1} \Lambda_{2}
\]
\begin{equation}\label{K24}
+ \Lambda_{2}^{*} S^{-1} \alpha \Lambda_{1} \Big)+8gf\Omega_{22} .
\end{equation}
Taking into account (\ref{K13}), (\ref{K19}) and
(\ref{K22})-(\ref{K24}) we derive
\begin{equation}\label{K25}
\wh v_t=gR_1+12gf\wh v+8gf\Omega_{22}^2.
\end{equation}
Finally, compare formulae (\ref{K18}), (\ref{K20}) and (\ref{K25})
to get
\[
\wh v_t-6gf\wh v=g\big(\wh v^{2}- \wh v_{xx}\big)_x.
\]
\end{proof}
\begin{Rk} 
To construct explicit solutions, take into account $f_t=12gf^2$
and notice that the matrix function $\a$ of the form
$\a(x,t)=f(t)(xI_n+a)$, where $a$ is an $n \times n$ matrix,
satisfies (\ref{Kd1}). It follows from (\ref{5}) and can be
also checked directly that
\[
\frac{\p^2}{\p x\p t}(\Lambda_s)=\frac{\p^2}{\p t\p x}(\Lambda_s) \quad (s=1,2).
\]
Thus, systems (\ref{K8}) and (\ref{K8'}) are compatible.
Using these systems and expression for $\a$, we can recover $\Lambda$ from the relations
\begin{equation}\label{K26}
\left[ \begin{array}{c}\Lambda_1(0,t)
\\ \Lambda_2(0,t) \end{array} \right]=\exp\{\b(t)\}
\left[ \begin{array}{c}\Lambda_1(0,0)
\\ \Lambda_2(0,0) \end{array} \right], 
\end{equation}
\begin{equation}\label{K26'}
\left[ \begin{array}{c}\Lambda_1(x,t)
\\ \Lambda_2(x,t) \end{array} \right]=\exp\{\g(x,t)\}
\left[ \begin{array}{c}\Lambda_1(0,t)
\\ \Lambda_2(0,t) \end{array} \right],
\end{equation}
where
\begin{equation}\label{K27}
\b(t)=\frac{1}{3}\big(f(t)-f(0)\big)
\left[ \begin{array}{lr}0 & a^2
\\ 0 & 0 \end{array} \right]+2\Big(\int_0^tg(u)f(u)du\Big)\left[ \begin{array}{lr}I_p & 0
\\ -2a & -I_p \end{array} \right],
\end{equation}
\begin{equation}\label{K28}
\g(x,t)=
\left[ \begin{array}{lr}0 & \frac{1}{2}f(t)x(xI_p+2a)
\\ -xI_p & 0 \end{array} \right].
\end{equation}
Function $f$ can be recovered from $g$ by the formula
$f(t)=1/(-12\p t^{-1}g)$.
Using (\ref{K12'}), (\ref{K13}) and (\ref{K26})-(\ref{K28}) one constructs $\wh v$
explicitly.
\end{Rk}
\section{Conclusion} \label{Conc}
\setcounter{equation}{0}
Thus, auxiliary systems for matrix nonisospectral
equations were introduced, and GBDT
version of the B\"acklund-Darboux transformation was
applied. It proved fruitful for the construction
of the explicit solutions of the nonisospectral
equations, including matrix equations and
equations with variable coefficients.
In our next work we plan to consider examples
with non-diagonal parameter matrices $A(x,t)$
in greater detail.

{\bf Acknowledgement.} \\ The work was supported by the Austrian
Science Fund (FWF) under Grant  no. Y330.

\end{document}